\documentclass[aps,twocolumn,pra,superscriptaddress]{revtex4-2}
\usepackage{amsfonts}
\usepackage[final,hiresbb]{graphicx}
\usepackage{amsmath}
\usepackage{amssymb}
\usepackage{amstext}
\usepackage{color}
\usepackage{braket}
\usepackage{leftindex}

\usepackage{subfigure}
\usepackage{appendix}
\usepackage{bbm}
\usepackage{bbold}

\definecolor{Green}{RGB}{0,204,102}
\definecolor{Purple}{RGB}{102,0,255}
\definecolor{Blue}{RGB}{51,153,255}
\definecolor{Red}{RGB}{255,010,010}

\newcommand{\PME}{\mathbb{P}_{\rm \! ME}}
\newcommand{\Pprod}{\mathbb{P}_{\rm \! prod}}
\newcommand{\psiprod}{\ket{\psi_{\rm {prod}}}}
\newcommand{\Pprodab}{\mathbb{P}_{\rm \! prod}^{\scriptscriptstyle \alpha\beta}}
\newcommand{\Pprodba}{\mathbb{P}_{\rm \! prod}^{\scriptscriptstyle \beta\alpha}}

\newcommand{\PTT}{\mathbb{P}_{\!\! {\scriptscriptstyle TT}}}
\newcommand{\Paa}{\mathbb{P}_{\!\! {\scriptscriptstyle \alpha\alpha}}}
\newcommand{\Pab}{\mathbb{P}_{\!\! {\scriptscriptstyle \alpha\beta}}}
\newcommand{\Pba}{\mathbb{P}_{\!\! {\scriptscriptstyle \beta\alpha}}}
\newcommand{\Pbb}{\mathbb{P}_{\!\! {\scriptscriptstyle \beta\beta}}}
\newcommand{\HE}{{\cal H}_{\! E}}
\newcommand{\psiME}{\psi_{{\rm\scriptscriptstyle ME}}}
\newcommand{\psiMEi}{\psi_{{\rm\scriptscriptstyle ME,i}}}
\newcommand{\psiMEf}{\psi_{{\rm\scriptscriptstyle ME,f}}} 
\newcommand{\psiMEfsolo}{\psi^{\rm \scriptscriptstyle solo}_{{\rm\scriptscriptstyle ME,f}}}

\newcommand{\Psolo}{\mathbb{P}_{\!\! \rm \scriptscriptstyle ME, solo}}
\newcommand{\Pavg}{\mathbb{P}_{\! \rm \scriptscriptstyle prod, solo}}
\newcommand{\HEsolo}{{\cal H}_{\! E, {\rm\scriptscriptstyle solo}}}

\begin{document}
\title{Entanglement Holonomy for Photon Pairs in Curved Spacetime}
	
\author{Mark T. Lusk}
\email{mlusk@mines.edu}
\affiliation{Department of Physics, Colorado School of Mines, Golden, CO 80401, USA}

\begin{abstract}
Polarization holonomy is analytically determined for maximally entangled photon pairs that transit a class of closed trajectories in the Kerr metric. This is used to define and investigate an \emph{entanglement holonomy} not associated with constituent product states.
\end{abstract}
\maketitle

\section{Introduction}

Since the seminal work of Pancharatnam almost seventy years ago\cite{Pancharatnam_1956}, electromagnetic radiation has provided an ideal setting for the exploration of holonomy. Experimentally explored at the classical level in optical fibers\cite{Tomita_1986}, polarization holonomy was subsequently replicated with single-photon measurements\cite{Kwiat_1991, Mair_2001, Voitiv_2024}. This makes sense since holonomies are a geometric characteristic of the domains themselves. 

The manifold to be probed is often a parameter space, such as the Poincar{\'e} Sphere\cite{Galvez_2003} or the Sphere of Modes\cite{Padgett_1999, Lusk_2022}, but it can also be spacetime.  When the spacetime metric is stationary, the polarization of light exhibits a helicity-dependent evolution that results in the rotation of linearly polarized planewaves\cite{Plebanski_1960, ThorneMisnerWheeler, FrolovShoom2011, Oancea_2020}. This Gravitational Faraday Rotation can be measured as the change in polarization between two observers\cite{FrolovShoom2011, Farooqui_2014}. In a recent work, it was shown that this effect can produce a projective holonomy provided the trajectory is such that outgoing and incoming polarizations can be measured at a common position\cite{Lusk_2024}---i.e. that the light traverses a closed circuit in three-space.

As in the laboratory setting, consideration of spacetime polarization holonomies can be extended to quantized Maxwell fields. The intent is not to seek new information about the metric but, rather, to understand how entanglement and curvature collectively influence the way in which holonomy is manifested. This has parallels with the rich field of holonomic quantum computing, which seeks to harness the geometry of a parameter space to encode information with increased resilience to noise\cite{Pachos_2000, Pinske_2023}. There an understanding of the relationship between entanglement and holonomy is particularly important for the design of quantum gates with multiple qubits\cite{Sjoqvist_2000}.  Entanglement holonomy in curved spacetimes should likewise play an important role in Relativistic Quantum Information Theory\cite{Mann_2012} and, more fundamentally, investigations of quantum physics in space\cite{Belenchia_2022}.

The present work focuses on the holonomy of photon pairs that follow closed circuits around a Kerr black hole.  Their maximally entangled initial state can be fully described by a single dyad orientation relative to a canonical configuration for which only one photon exhibits holonomy. A purely analytical approach is used, and the expressions developed are compared with the holonomy accumulated by the same photons in product states. The difference is dubbed an \emph{entanglement holonomy}, attributable to the projection of the initial state of one mode onto the final state of another. It is a signature feature of entanglement that such projections exist. The new holonomy is quantified and interpreted for a family of closed circuits within a spherical surface in Boyer-Lindquist coordinates. Influences of black-hole rotation rate, initial dyad orientation, and spacetime starting point are elucidated.

\section{Canonical Polarization Holonomy}

We first review polarization holonomy within the classical setting, immediately applicable to single-photon states as well, to establish a foundation for the consideration of entanglement. Whether quantized or not, the vacuum propagation of light is assumed to be governed by the covariant form of a Maxwell Lagrangian\cite{ThorneMisnerWheeler}. Plane-wave dynamics can be subsequently re-cast as a set of ordinary differential equations provided the wavelength is much shorter than any characteristic lengths---the geometric optics approximation\cite{ThorneMisnerWheeler}. At higher order, there is a helicity dependence to photon trajectories, a gravitational Spin Hall effect\cite{Oancea_2020}, but this is not considered in the present work.  Within the Kerr metric, it is then possible to analytically construct trajectories that are closed in three-space and to track the evolution of polarization\cite{Gralla_2020, Wang_2022}. Each trajectory is characterized by tangent vectors that are discontinuous  at the common location of source and receiver.  A polarization misorientation angle, $\chi$, is obtained from the projection of initial and final vectors: 
\begin{equation}\label{PTT}
\PTT := \braket{T_f|T_i} \equiv \cos\chi .
\end{equation}
Here  $\mathbb P$ is used to indicate the result of a projection operation, while subscripted pairs identify the final and initial states, in that order. This polarization holonomy was recently explored for a class of spherically-confined circuits\cite{Lusk_2024}. 

\begin{figure}[t]
	\begin{center}
		\includegraphics[width=0.75\linewidth]{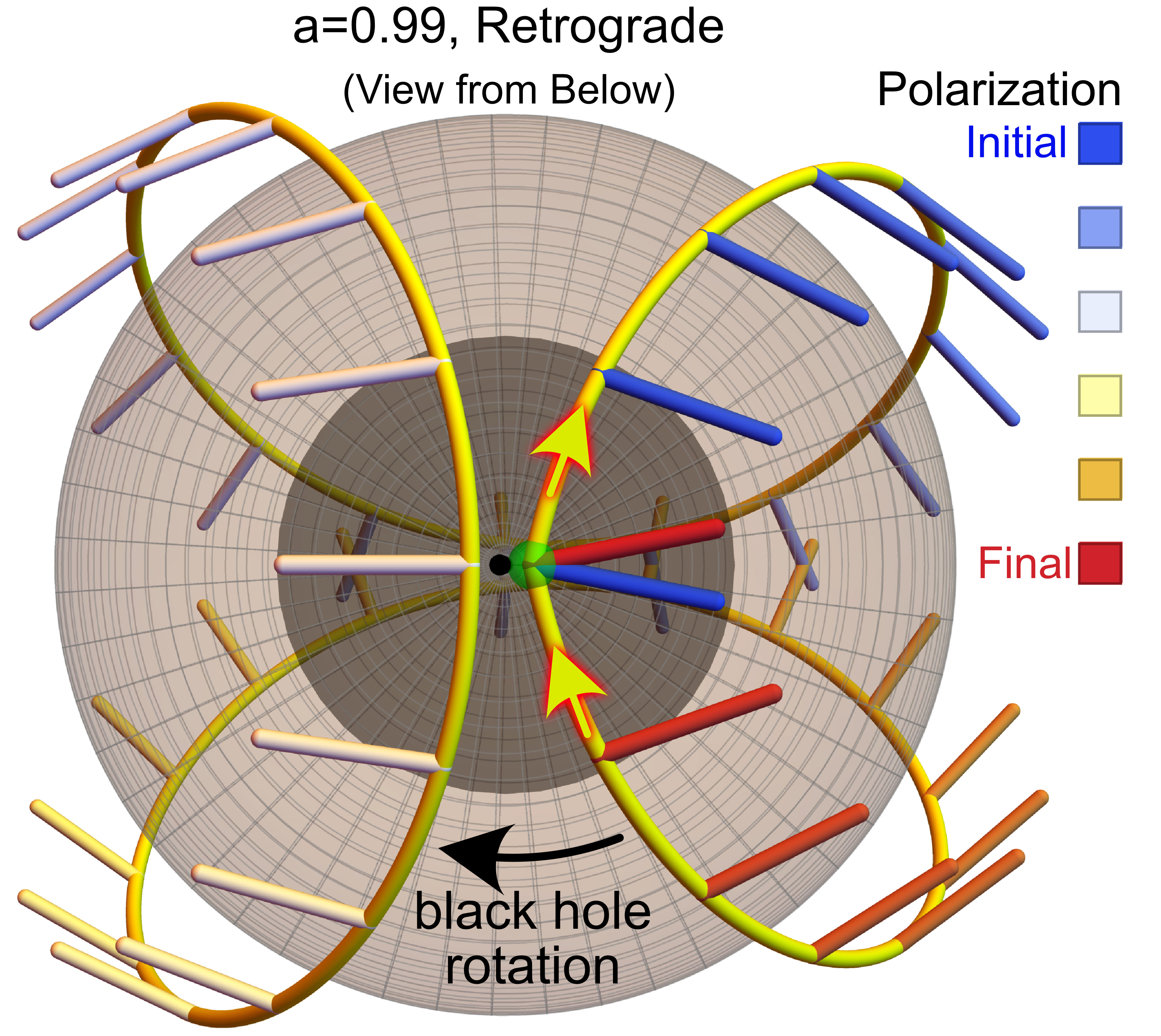}
	\end{center}
	\caption{ \emph{Evolution of Polarization}. The 3-space projection of a retrograde trajectory is shown in yellow for dimensionless black-hole rotation rate $a=0.99$ and start/end polar angle $\theta_0=177.3^\circ$. Along this path, the polarization of light is plotted as colored sticks starting with blue and ending red. The view is from below. Since the temporal component is engineered to always be zero, the angle between the red and blue sticks is the polarization holonomy, $\chi$. The figure includes the outer ergosphere (light gray) and the outer event horizon (dark gray).} 
	\label{a0p99_retro_polarization_below}
\end{figure}

\begin{figure}[t]
	\begin{center}
		\includegraphics[width=0.75\linewidth]{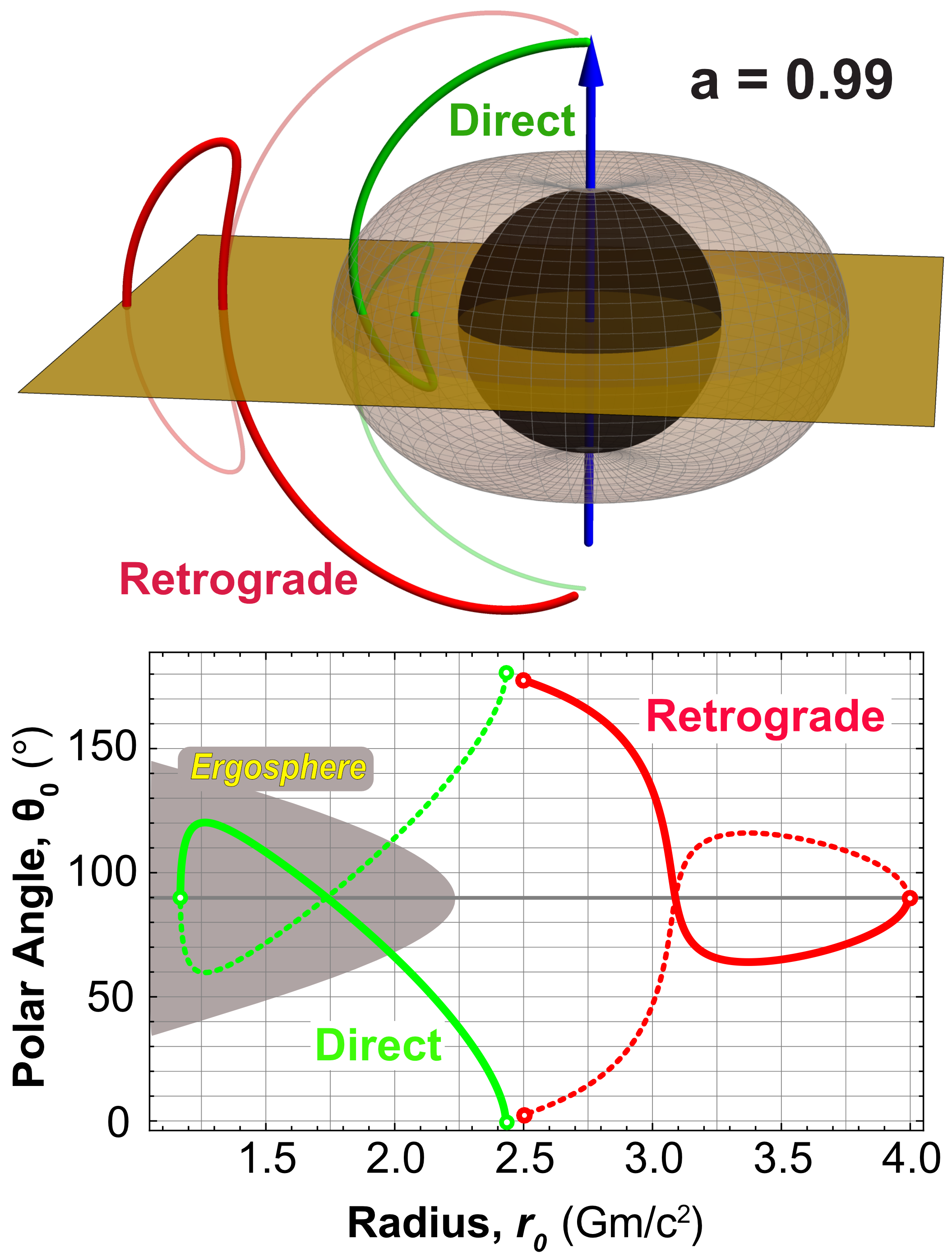}
	\end{center}
	\caption{ \emph{Send/receive points resulting in closed circuits.} Sets of retrograde (red) and direct (green) start/end points are shown for dimensionless black-hole rotation rate $a = 0.99$. Green and red curves that are semi-transparent (top) and dashed (bottom) identify admissible complementary sets ($\theta\rightarrow 180^\circ - \theta$). The top plot shows a 3-space projection of the positions, with the outer ergosphere shown in gray and the outer event horizon in black. The choice of azimuthal orientation is arbitrary. Note that, for the direct case, it is possible to start/end within the ergosphere even though the light may escape it before returning.} 
	\label{start_end_set_a0p99}
\end{figure}

A given trajectory can support two orthogonal polarization vectors. Interestingly though, it is possible to construct initial polarization states such that only one vector exhibits holonomy. In particular, polarizations (R) that are initially parallel to the local radial coordinate line, never exhibit  holonomy. Because of this, we need only construct the initial polarization (T) to be tangent to the spherical surface and orthogonal to the trajectory. One such trajectory is shown in Fig. \ref{a0p99_retro_polarization_below} along with the evolution of its T-polarization vector. The angle between the initial (blue) and final (red) polarizations is the holonomy, identified in this work as $\chi$.  A class of such trajectories was identified, and the start/end points are shown in Fig. \ref{start_end_set_a0p99}.

%
%
%

%
%
%

\begin{figure}[t]
	\begin{center}
		\includegraphics[width=1.0\linewidth]{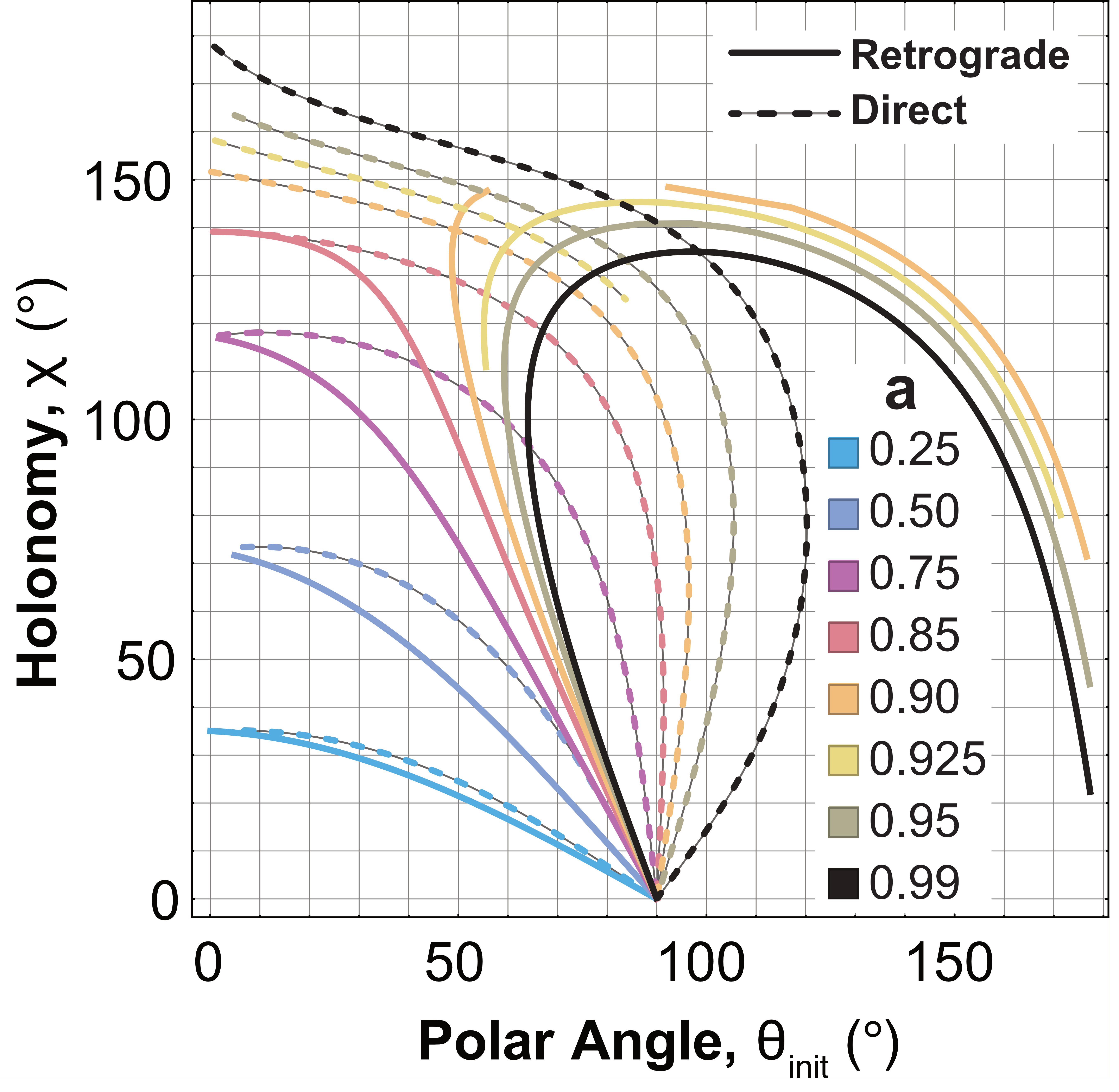}
	\end{center}
	\caption{ \emph{Polarization Holonomy}. The angle, $\chi$, between initial and final polarizations as a function of black-hole rotation rate, $a$, and start/end polar angle, $\theta_0$. Results for retrograde trajectories are shown with solid curves while results for direct trajectories are dashed curves with a light gray outline to guide the eye.} 
	\label{Holonomy_Comparisons}
\end{figure}

For each circuit, an associated polarization holonomy was calculated analytically, and the results are plotted in Fig. \ref{Holonomy_Comparisons} as a function of the polar angle, $\theta_0$, and singularity rotation rate, $a$.

%
%
%

The results just summarized are associated with classical electromagnetic fields, but they apply equally well to the transit of single photons considered using quantum field theory in curved spacetime\cite{Winitzki_2007, Bar_2019}. While entanglement is frame dependent in general, the current setting obviates the need to consider this because photon pairs are both produced and eventually measured within a single, stationary frame\cite{Fuentes_2005}. The relevant measurements can therefore be couched within the quantum field theory of Minkowski spacetime.

\section{Holonomies for a Rotated Polarization Basis}

Our current investigation builds on single-photon dynamics by considering the nature of holonomy associated with photon pairs that are maximally entangled in polarization\cite{Kwiat_1995}. At issue is the influence that entanglement has on holonomic manifestations.  Instead of considering tangential and radial polarizations $\ket{T}$ and $\ket{R}$, though, it will prove useful to define more general single-photon modes constructed by rotating this dyad about the initial direction of propagation. To that end, define a dyad rotation matrix, $\Omega(\gamma)$:
\begin{equation}\label{Omega}
\Omega(\gamma) = \left(
\begin{array}{cc}
\cos\gamma & \sin\gamma \\
-\sin\gamma & \cos\gamma  \\
\end{array}
\right).
\end{equation}
New initial polarizations, $\ket{\alpha}$ and $\ket{\beta}$, are then defined as
\begin{equation}\label{dyad}
\left(
\begin{array}{c}
\ket{\alpha_i}\\
\ket{\beta_i}\\
\end{array}
\right) = \Omega(\gamma) \left(
\begin{array}{c}
\ket{T_i}\\
\ket{R}\\
\end{array}
\right) ,
\end{equation}
so that
\begin{align}\label{alphabeta}
\ket{\alpha_i} &= \cos\gamma\ket{T_i} + \sin\gamma\ket{R}\nonumber \\
\ket{\beta_i} &= -\sin\gamma\ket{T_i} + \cos\gamma\ket{R}.
\end{align}
Since $\braket{\alpha_i | \beta_i} = 0$, the new vectors comprise an orthogonal, $\gamma$-dependent polarization basis.

\begin{figure}[t]
	\begin{center}
		\includegraphics[width=0.95\linewidth]{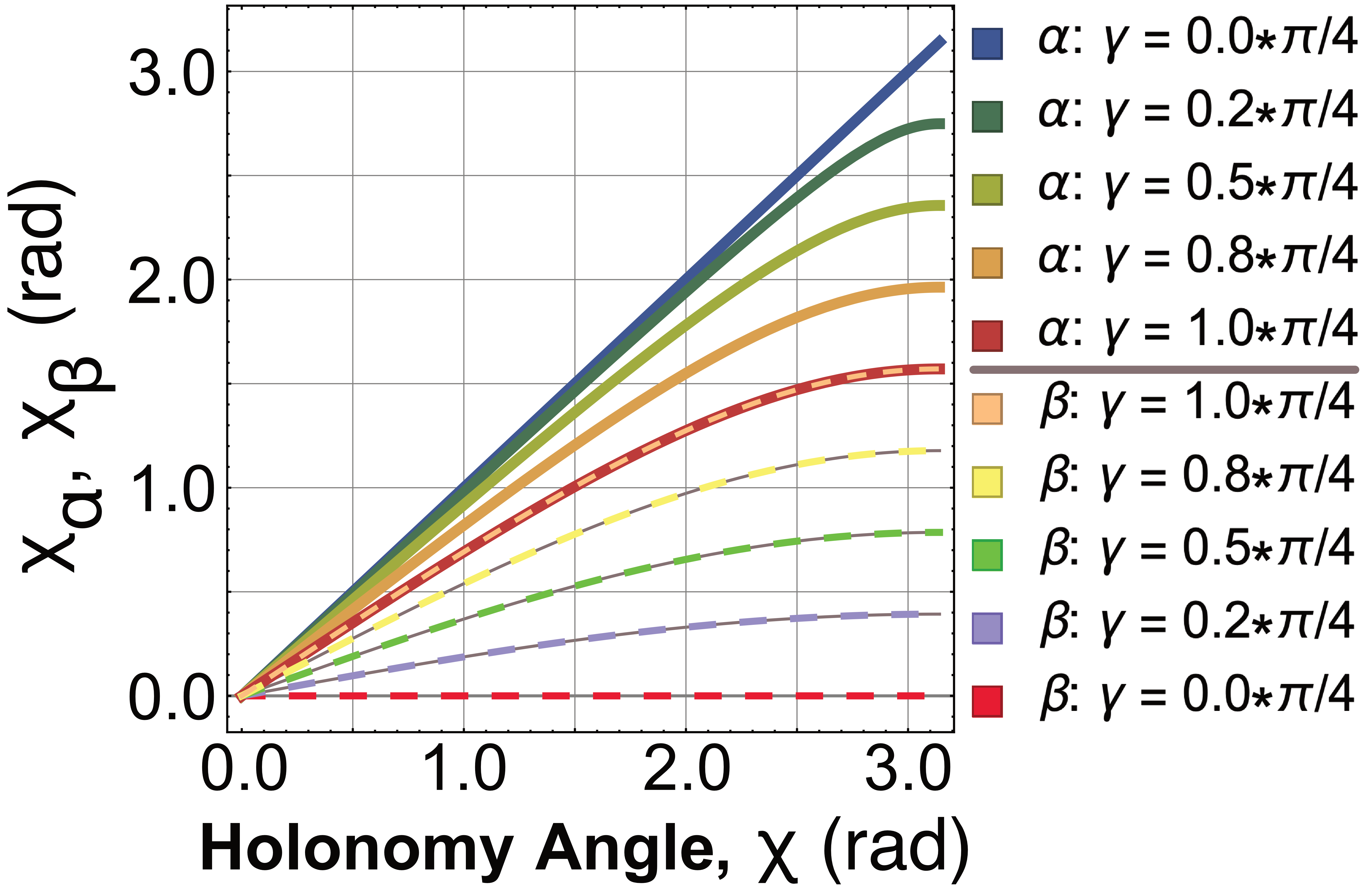}
	\end{center}
	\caption{ \emph{Holonomy Angles for Single Photons.} Holonomy angles $\chi_\alpha$ and $\chi_\beta$ are plotted as functions of T-polarization holonomy angle, $\chi$, and dyad rotation angle, $\gamma$. Results for $\alpha$-mode photon 1 are shown with solid curves while those for $\beta$-mode photon 2 are dashed curves with a light gray outline to guide the eye.} 
	\label{Holonomy_Angles_Separated_Photons}
\end{figure}

Suppose that unentangled photons 1 and 2 are constructed in modes $\ket{\alpha_i}$ and $\ket{\beta_i}$, respectively, and sent along the same path that produced projection $\PTT$ of Eq. \ref{PTT}. Holonomies are then associated with each photon, $\Paa = \braket{\alpha_f | \alpha_i}$ and $\Pbb = \braket{\beta_f | \beta_i}$. Recalling that $\braket{T_i | R} = \braket{T_f | R} = 0$, these evaluate to
\begin{align}\label{PaaPbb}
\Paa &\equiv \cos\chi_\alpha = \cos^2\gamma \cos\chi + \sin^2\gamma\\ \nonumber
\Pbb &\equiv \cos\chi_\beta = \sin^2\gamma \cos\chi + \cos^2\gamma .
\end{align}
The holonomy angles, $\chi_\alpha$ and $\chi_\beta$, are expressed with respect to the T-polarization holonomy angle, $\chi$. This makes for a convenient scaffold from which to study how basis rotation influences misorientation measurement. The holonomy angles are plotted in Fig. \ref{Holonomy_Angles_Separated_Photons}. The two lines associated with $\gamma=0$ are equivalent to the case considered previously. We can always associate this with a particular trajectory by specifying the starting position of the photon path, Fig. \ref{Holonomy_Comparisons}, reading off the holonomy angle, $\chi$, and then applying the value to Fig. \ref{Holonomy_Angles_Separated_Photons}. 

%
%
%

The projections of Eq. \ref{PaaPbb} can be applied to quantify holonomy for two-photon product state
\begin{equation}\label{sep} 
\psiprod:= \ket{\alpha}_1 \ket{\beta}_2 ,
\end{equation}
for which the associated projection is
\begin{equation}\label{Pprod}
\Pprodab := \braket{ \psi_{\rm {prod}} | \psi_{\rm {prod}}} = \braket{\alpha_f | \alpha_i}_1 \braket{\beta_f | \beta_i}_2 = \Paa\Pbb.
\end{equation}
Of course, a second two-photon state, $\psiprod := \ket{\beta}_1 \ket{\alpha}_2$, will produce projection $\Pprodba \equiv \Pprodab$, and these will both be referred to as $\Pprod$. With the projections of these product states in mind, we can now turn to the question of how entanglement might affect holonomy.

\section{Entangled State Holonomy}

The holonomies of constituent product states can be used to quantify their maximally entangled (ME) counterpart,
\begin{equation}\label{ME}
\ket{\psiME} := \frac{1}{\sqrt{2}}\left( \ket{\alpha}_1 \ket{\beta}_2 + \ket{\beta}_1 \ket{\alpha}_2 \right).
\end{equation}
Such bi-photons can also be sent through a holonomy-producing circuit and projected onto their initial state in the originating, stationary laboratory frame. As detailed in Appendix A, rotation of the initial polarization dyad results in inequivalent entangled states. It is further shown, in Appendix B, that maximal entanglement is preserved in the final state. 

\begin{figure}[t]
	\begin{center}
		\includegraphics[width=0.85\linewidth]{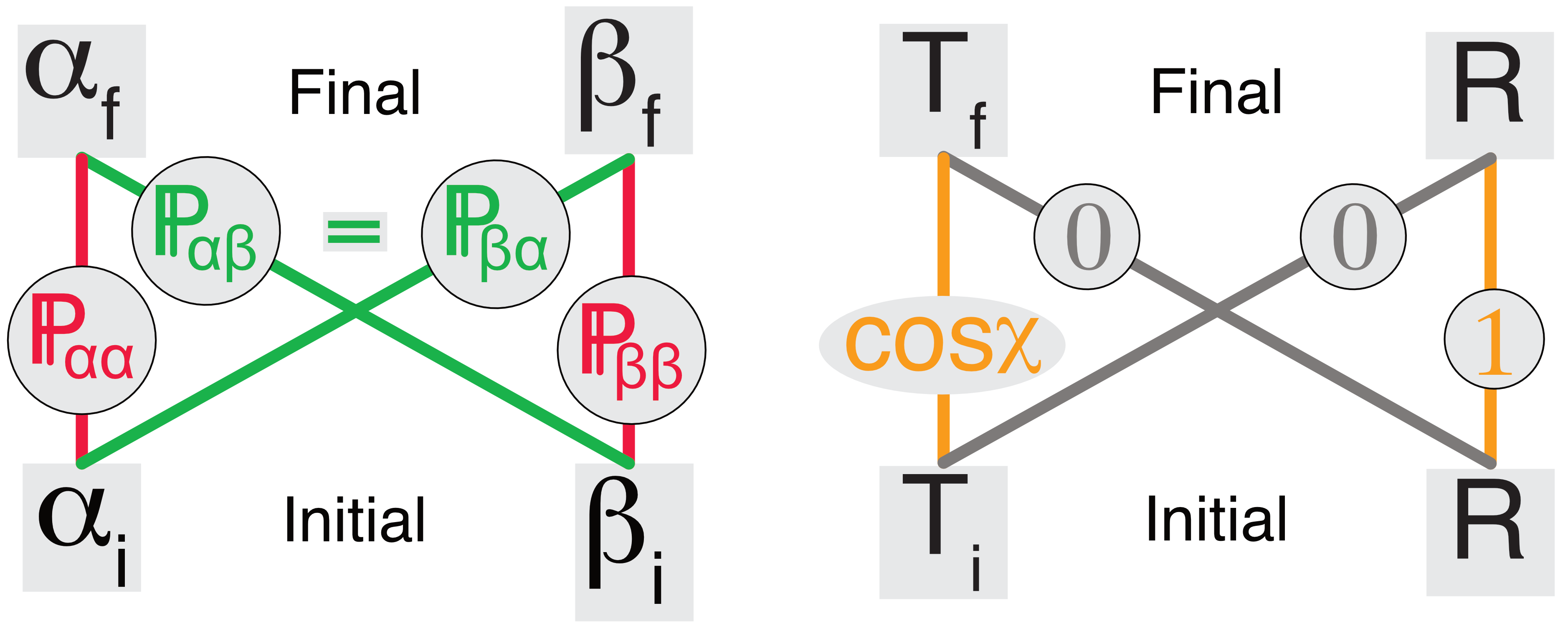}
	\end{center}
	\caption{ \emph{Entanglement Projections.} (Left) Schematic diagram of the four projections used to construct $\PME$ of Eq. \ref{PME1}. The red projections are those of the product state, while the green projections are new contributions that are attributable to entanglement. (Right) Recalling the form of $\ket{\alpha}$ and $\ket{\beta}$ from Eq. \ref{alphabeta}, the schematic links the four projections at left to the T-polarization holonomy angle, $\chi$.} 
	\label{Entanglement_Projection_Schematic_2}
\end{figure}

The resulting projection evaluates to a combination of four single-photon projections,
\begin{equation}\label{PME1}
\PME = \braket{ \psiMEf | \psiMEi } = \Paa\Pbb + \Pab\Pba ,
\end{equation}
and their relationships are given schematically in the left panel of Fig. \ref{Entanglement_Projection_Schematic_2}. The red projections are those of the product state, while the green projections are new contributions that are attributable to entanglement. Recalling the expressions for $\ket{\alpha}$ and $\ket{\beta}$ of Eq. \ref{alphabeta}, the right panel of Fig. \ref{Entanglement_Projection_Schematic_2} links all four projections to the T-polarization holonomy angle, $\chi$. 

%
\begin{figure}[t]
	\begin{center}
		\includegraphics[width=0.90\linewidth]{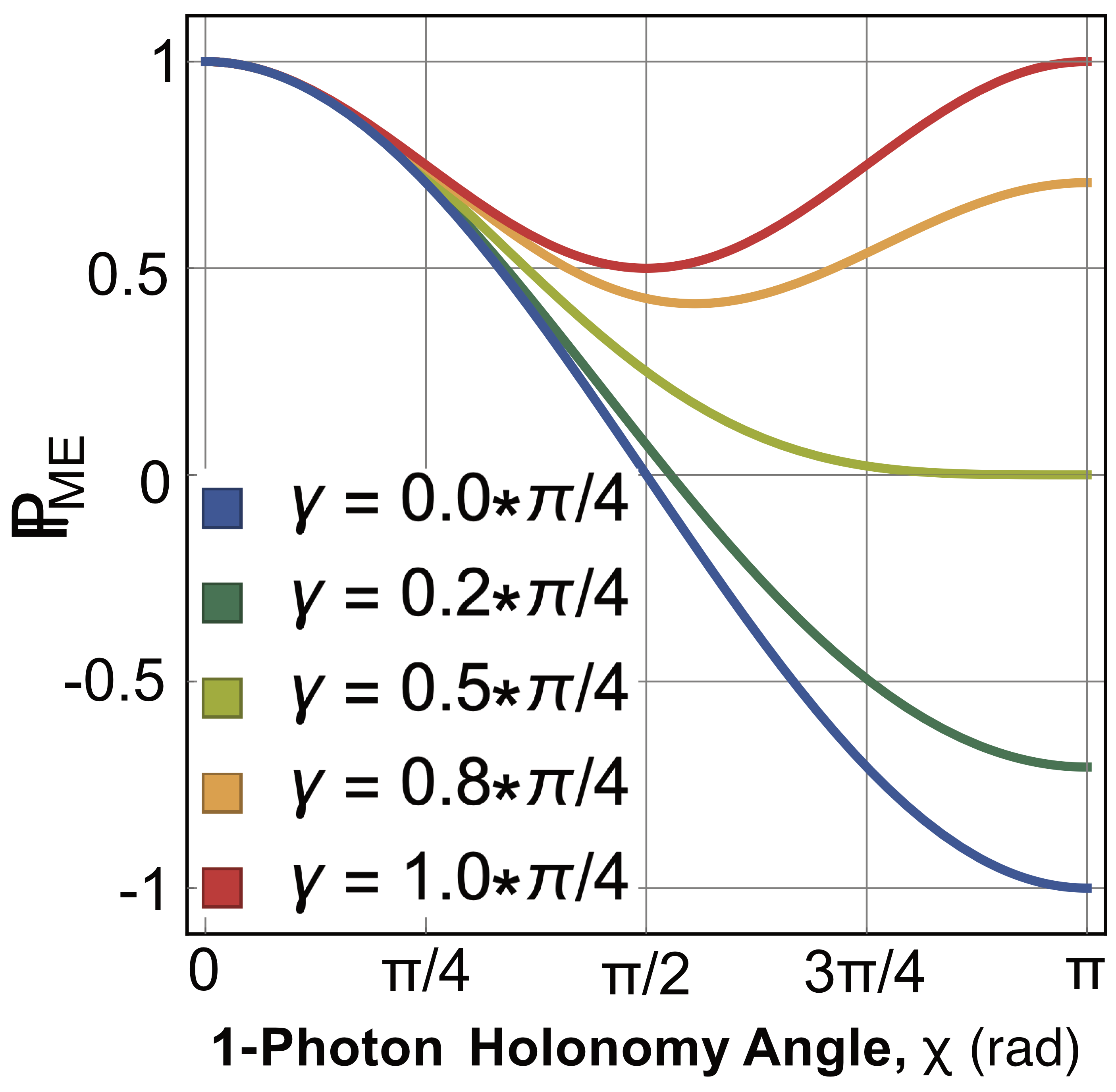}
	\end{center}
	\caption{ \emph{Entangled State Projection, $\PME$.} The projection of a maximally entangled state, $\PME$ Eq. \ref{PME2}, is plotted as a function of T-polarization holonomy angle, $\chi$, and dyad rotation angle,  $\gamma$. For $\chi=0$, the initial and final states are the same, so $\PME=1$. Other extremes and trends are discussed in the text.} 
	\label{Projection_Max_Entangled}
\end{figure}

Substitution of the individual component expressions yields the projection of $\ket{\psiMEf}$ onto $\ket{\psiMEi}$:
\begin{equation}\label{PME2}
\PME = \cos^2(2\gamma) \cos\chi + \frac{1}{4}\sin^2(2\gamma) \left(3 + \cos(2\chi) \right) .
\end{equation}
This is plotted in Fig. \ref{Projection_Max_Entangled} over a range of dyad rotation angles, $\gamma$. The projection ranges between $+1$ and $-1$ and so could be expressed in terms of an angle. Its geometric interpretation is not immediately evident though. An analysis of the trends observable in Fig. \ref{Projection_Max_Entangled} are put off until they are compared directly with product states below.

%
%
%

\section{Entanglement Holonomy}

We have constructed contrasting projections associated with two-photon states: $\Pprod$ for product states and $\PME$ for maximally entangled photons. Effects due solely to entanglement can now be extracted by taking their difference, dubbed the entanglement holonomy:
\begin{equation}\label{HE1}
\HE := \PME - \frac{1}{2}(\Pprodab + \Pprodba ) = \PME - \Pprod .
\end{equation}
The combination of Eqs. \ref{PME1} and \ref{Pprod} implies that $\PME = \Pprod + \Pab\Pba$. Therefore
\begin{equation}\label{HE2}
\HE = \Pab\Pba = \sin^2(2\gamma) \sin^4\left(\frac{\chi}{2}\right) .
\end{equation}

\begin{figure}[t]
	\begin{center}
		\includegraphics[width=0.95\linewidth]{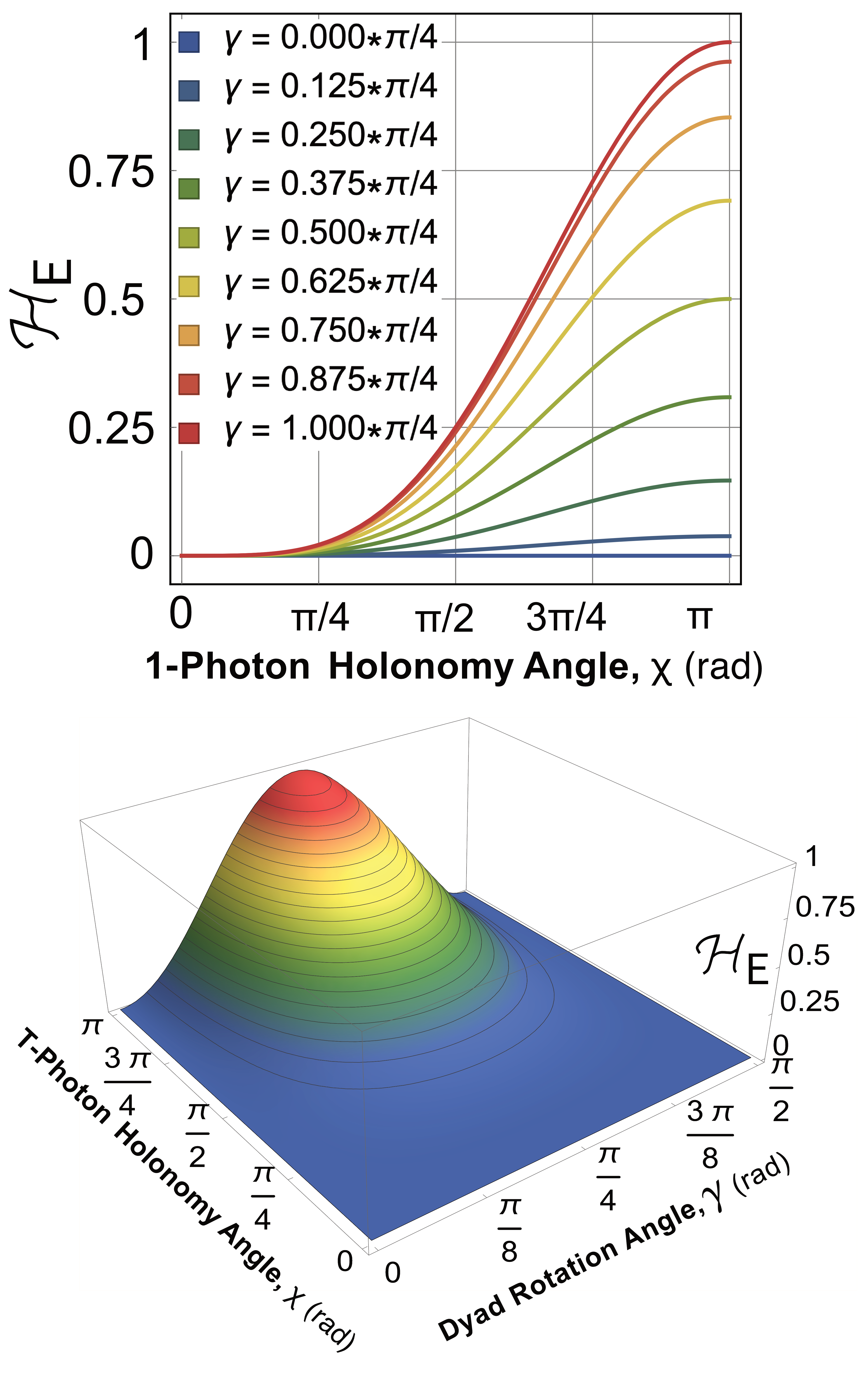}
	\end{center}
	\caption{ \emph{Entanglement Holonomy, $\HE$.} The difference in the projections of maximally entangled states, $\PME$ of Eq. \ref{PME2}, and product states, $\Pprod$ of Eq. \ref{Pprod} is a measure of the holonomy uniquely attributable to photon entanglement, Eq. \ref{HE1}. This is plotted as a function of T-polarization holonomy angle, $\chi$, and dyad rotation angle, $\gamma$. The surface plot in the bottom panel covers twice the range of dyad rotation angle to give a clearer visualization of $\HE$.} 
\label{Entanglement_Holonomy}
\end{figure}

Entanglement holonomy arises from the projection of the initial state of one mode onto the final state of another. It is a signature feature of entanglement that these projections exist within the overlap of initial and final two-photon states. The holonomy is plotted in Fig. \ref{Entanglement_Holonomy} over a range of values of T-polarization holonomy angles, $\chi$, and dyad rotation angles, $\gamma$.

%
%
%

The figure shows that there is no entanglement holonomy if the dyad basis has not been rotated ($\gamma = 0$) despite  projection $\PME$ ranging between $\pm 1$. This is because there is no projection between polarization vectors $\ket{T}$ and $\ket{R}$.  As dyad rotation angle $\gamma$ increases, $\HE$ exhibits an increasingly strong dependence on the single-photon holonomy angle, $\chi$.  This implies that rotation of the polarization dyad has a monotonic influence on $\HE$. 


The maximum entanglement holonomy is associated with initial dyad vectors that have been rotated by $\gamma = 90^\circ$. This evenly distributes the holonomy-generating polarization, $\ket{T}$:
\begin{align}\label{max1}
\bra{\alpha_f} &= \frac{1}{\sqrt{2}} \left( \bra{T_f} + \bra{R} \right) \nonumber \\
\ket{\beta_i} &= \frac{1}{\sqrt{2}} \left( -\ket{T_i} + \ket{R} \right) .
\end{align}
The result is that cross-term projection $\Pab$ is large because the $\alpha$-mode rotates $90^\circ$ and becomes aligned with the initial $\beta$-mode. Likewise, the $\beta$-mode rotates $90^\circ$ and ends up aligned with initial $\alpha$-mode: $\Pab=\Pba=1$. This further implies that $\Paa=\Pbb=0$ and that the product state projection, $\Pprod$, of Eq. \ref{Pprod} is zero as well.  With the initial dyad thus set, consider a trajectory would rotate a T-mode photon by $\chi\!=\!\pi$ so that $\ket{T_f}\!=\!-\ket{T_i}$.  The projection of the maximally entangled state, Eq. \ref{PME2}, is then equal to one, as is clear in the red curve of Fig. \ref{Projection_Max_Entangled}. This explains why the entanglement holonomy of Fig. \ref{Entanglement_Holonomy} is maximized for $\gamma=\pi/4$ and $\chi=\pi$. The  projection of the final state onto the initial state exhibits a maximum in the cross-term projections at the same time that the modes themselves rotate so as to be orthogonal to their initial state, and $\HE\!=\!1$.

\section{Entangled States with Solo Photon Transit}
 
Before applying our result to the Kerr system, it is worth considering the notion that an entanglement holonomy might be produced even if one of the photons is kept in the lab while the other traverses a circuit. The final state would then be
\begin{equation}\label{MEsolo}
\psiMEfsolo := \frac{1}{\sqrt{2}}\left( \ket{\alpha_f}_1 \ket{\beta_i}_2 + \ket{\beta_f}_1 \ket{\alpha_i}_2 \right),
\end{equation}
and the projection of initial and final states reduces to
\begin{equation}\label{PMEsolo}
\Psolo = \frac{1}{2} \left(\Paa+\Pbb\right). 
\end{equation}

To calculate the associated entanglement holonomy, an appropriate projection must be identified to replace $\Pprod$ of Eq. \ref{Pprod}. Since single-photon transits generate a projection of either $\Paa$ or $\Pbb$, while the photon kept in the lab always has a projection of 1, the appropriate product state projection is
\begin{equation}\label{Pavg}
\Pavg := \frac{1}{2} \left(\Paa*1 + \Pbb*1 \right).
\end{equation}
The entanglement holonomy is the difference between $\Psolo$ and $\Pavg$:
\begin{equation}\label{HEsolo}
\HEsolo = \Psolo - \Pavg = 0.
\end{equation}
There is no entanglement holonomy in this case. Both photons must traverse the circuit to produce a projection attributable to entanglement. 

\section{Application to Kerr Circuits}

\begin{figure}[t]
	\begin{center}
		\includegraphics[width=1.0\linewidth]{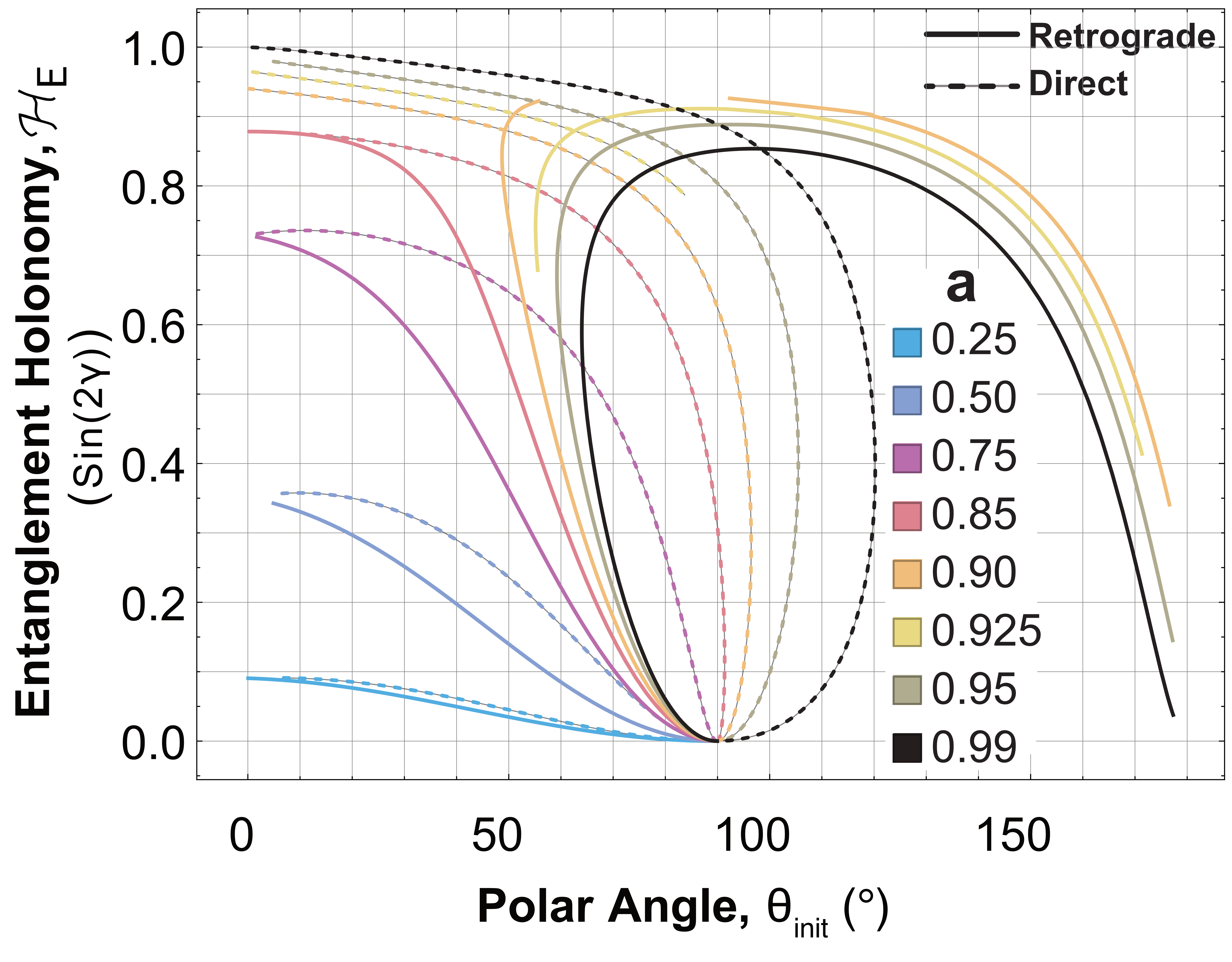}
	\end{center}
	\caption{ \emph{Polarization Holonomy}. The entanglement holonomy, ${\cal H}_E$, between initial and final polarizations as a function of black-hole rotation rate, $a$, and start/end polar angle, $\theta_0$. The entanglement holonomy is normalized by plotting it in units of $\sin(2\gamma)$ as this removes the explicit dependence on the dyad rotation angle, $\gamma$.  Results for retrograde trajectories are shown with solid curves while results for direct trajectories are dashed curves with a light gray outline to guide the eye.} 
	\label{Entangled_Holonomy_Comparisons}
\end{figure}

With the construction of an entanglement holonomy in hand, we can now return to the original class of closed circuits on spherical surfaces around a Kerr black hole. Each T-polarization holonomy in Fig. \ref{Holonomy_Comparisons} can be mapped to a corresponding value of $\HE$ using Eq. \ref{HE2}, and the resulting family of entanglement holonomies is presented in Fig. \ref{Entangled_Holonomy_Comparisons}. As is clear from Eq. \ref{HE2}, the holonomy dependence on dyad rotation angle, $\gamma$, is distilled into a coefficient, $\sin(2\gamma)$. It is therefore chosen as the unit of measurement of $\HE$ in the plot. A given value of $\HE$ will increase monotonically with $\gamma$ as the dyad rotation increases to $\pi/4$.

For starting polar angles sufficient close to the north pole, $\HE$ increases monotonically with black-hole rotation rate, $a$, with the holonomy larger for direct orbits. The trend reverses for polar angles that are sufficiently close to the south pole.  There is a range of Boyer-Lindquist polar angles, $\theta_{\rm\scriptstyle init}$, for which there two values of $\HE$ are possible. This is because two initial radii are sometimes supported for a given polar angle, as shown in Fig. \ref{start_end_set_a0p99}. The largest entanglement holonomy is associated with direct orbits that start and end at the north pole. 

%
%
%
%

In the special case for which the black hole does not rotate ($a\!=\!0$), spacetime reduces to the Schwarzschild metric, and Fig. \ref{Entangled_Holonomy_Comparisons} makes it clear that the entanglement holonomy is then zero. This makes sense, as we have previously established that black-hole rotation is required to produce a polarization holonomy\cite{Lusk_2024}, just as it is to produce Gravitational Faraday Rotation\cite{Frolov_2011}. The result has simply been extended here to pairs of maximally entangled 2-photon states.

\begin{figure}[t]
	\begin{center}
		\includegraphics[width=0.65\linewidth]{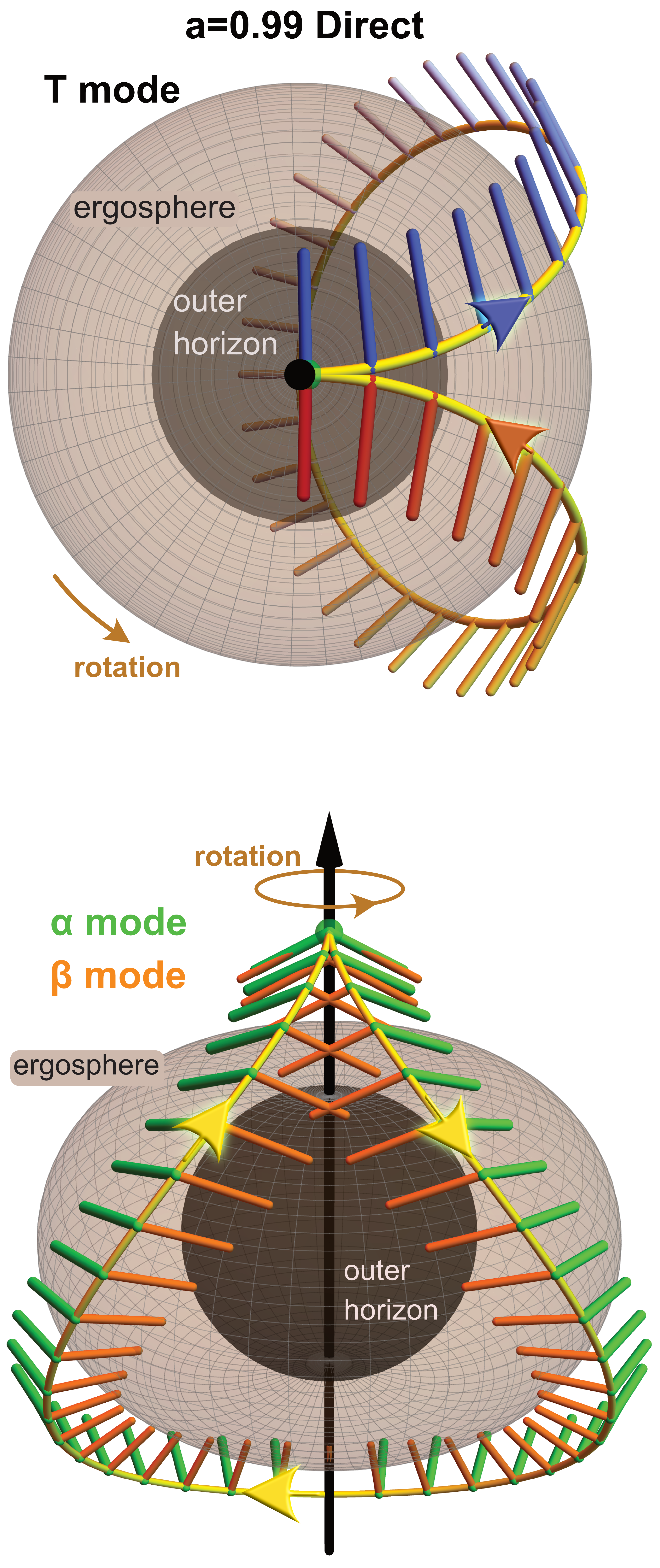}
	\end{center}
	\caption{ \emph{Kerr Trajectory with Maximum Entanglement Holonomy}. The 3-space projection of a direct trajectory is shown in yellow for dimensionless black-hole rotation rate $a=0.99$ and start/end polar angle $\theta_0=0.85^\circ$. The T-mode holonomy angle of $\chi=176.9^\circ$ and dyad rotation of $\gamma=\pi/4$ yield a value of entanglement holonomy of $\HE= 0.999$, very close to the maximum possible. (Top Panel) The T-polarization of light is plotted as colored sticks starting with blue and ending red. The view is from above. Since the temporal component is engineered to always be zero, the angle between the red and blue sticks is the polarization holonomy, $\chi$. (Bottom Panel) The evolving $\alpha$-polarization is plotted in green while the $\beta$-polarization is in orange. Both panels include the outer ergosphere (light gray) and the outer event horizon (dark gray).} 
	\label{traj_a0p99_direct}
\end{figure}

We have previously discussed the limiting case for which $\HE$ is a maximum: $\chi=\pi$ and $\gamma=\pi/4$. This is realized for the transit shown in Fig. \ref{traj_a0p99_direct}. The top panel shows the evolution of T-polarization, while the bottom panel allows the $\alpha$-mode and $\beta$-mode to be visualized. This second image shows how the projection of the final state of mode $\ket{\alpha} (\ket{\beta})$ is precisely the initial state of mode $\ket{\beta} (\ket{\alpha})$. This is what maximizes the entanglement holonomy.

%
%
%
%

\section{Discussion}

Pairs of entangled photons exhibit a polarization holonomy not attributable to that of either photon individually. Rather, it is due to the projection of an initial state of one mode onto the final state of another, a signature feature of entanglement and lack of local-realism\cite{Bell_1964, Kwiat_1995}. The result is consistent with the finding that no such holonomy is produced unless both photons traverse the singularity, and it has implications for the design of laboratory experiments that seek to make analogous measurements for geodesics of the relevant parameter space. Entanglement holonomy has been quantified for a family of closed circuits about a Kerr black hole. It is influenced by black-hole rotation rate, rotation direction, orientation of the initial polarization dyad, and spacetime starting point. 

A lowest-order geometric optics setting was employed, for which photon trajectories are not influenced by spin. This is not the case for higher-order approximations\cite{Frolov_2011, Oancea_2020}, which should not be an issue provided the returning photons can still be measured relative to the same stationary frame. Quantization of the electromagnetic field was carried out within the setting of quantum field theory since only a single frame is involved in the measurements. If the photon polarizations are to be measured in frames that have a relative acceleration, or if the sending and receiving frames have a relative acceleration, then the dynamics must be reconsidered using quantum field theory in curved spacetime. In the present work, the region considered is idealized as a vacuum to avoid complications associated with accretion disks or Hawking radiation. 

Using analytical expressions for closed-form Kerr trajectories\cite{Gralla_2020, Wang_2022}, it should be possible to identify a wide range of self-intersecting orbits that could form the basis for further exploring both single-photon and entangled-photon polarization holonomies. A second study, in which two maximally entangled photons traverse distinct closed circuits, is likely to also lend new insights. A third  extension of the current work would be to consider the possible influence of entanglement on Gravitational Faraday rotation---i.e. on open circuits. This could be carried out numerically using a combination of parallel transport for the trajectories and Fermi-Walker transport for the polarization dyad. The setting would allow for the study of an Aharonov-Bohm effect\cite{Aharonov_1959} associated with the distinct transits of entangled photons that start at a common point and end at a second common point. It should also be mentioned that, although entanglement was taken to be maximal for the sake of clarity here, the approach can be readily adapted to partially entangled states as well.


\appendix

\section{The Inequivalence of Rotated, Maximally Entangled States}

When considering Bell tests using polarization, it proves both insightful and useful to write the same two-photon state in two equivalent representations using distinct pairs of maximally entangled modes. The first is in terms of horizontal and vertical polarizations, $\ket{H}$ and $\ket{V}$, while the second is in terms of circular polarizations, $\ket{+}$ and $\ket{-}$. First write the state with respect to linear polarizations:
\begin{equation}
\ket{\Psi} = \frac{1}{\sqrt{2}} \left( \ket{H}\ket{V} + \ket{V}\ket{H} \right).
\end{equation}
Since
\begin{align}
\ket{H} &= \frac{1}{\sqrt{2}} ( \ket{-} - \ket{+} ) \nonumber \\
\ket{V} &= \frac{\imath}{\sqrt{2}} ( \ket{-} + \ket{+} ) ,
\end{align}
substitution into the original form of $\ket{\Psi}$ gives the following equivalent form:
\begin{equation}
\ket{\Psi} = \frac{\imath}{\sqrt{2}} \left( \ket{-}\ket{-} - \ket{+}\ket{+} \right) .
\end{equation}

The fact that two sets of maximally entangled basis vectors can be used to describe the same state may make it tempting to conclude that the canonical bi-photon state,
\begin{equation}
\ket{\psi_0} = \frac{1}{\sqrt{2}} \left( \ket{T}\ket{R} + \ket{R}\ket{T} \right), 
\end{equation}
is somehow equivalent to 
\begin{equation}
\ket{\psi(\gamma)} = \frac{1}{\sqrt{2}} \left( \ket{\alpha(\gamma)} \ket{\beta(\gamma)} + \ket{\beta(\gamma)} \ket{\alpha(\gamma)} \right) ,
\end{equation}
where $\ket{\alpha(\gamma)}$ and $\ket{\beta(\gamma)}$ are defined in Eq. \ref{dyad}. This is not true though. To see why, write out the second bi-photon state in terms of canonical basis, $\ket{T}$ and $\ket{R}$, then evaluate the inner product of this vector for two distinct values of dyad rotation to find that
\begin{equation}
\braket{\Psi(\gamma_1) | \Psi(\gamma_2) } = \cos(\gamma_1 - \gamma_2).
\end{equation}
A necessary condition for equivalence is that this inner product be equal to one, but this is only the case when the two bases are the same---i.e. when $\gamma_1 = \gamma_2$. Comparison with the canonical state amounts to a special case for which $\gamma_1 = 0$. This proves that entangled states constructed from a rotated dyad are not equivalent to entangled states composed of the basis vectors of the unrotated dyad.

\section{Proof That the Final State is Maximally Entangled}

Bi-photons can be sent through a holonomy-producing circuit and projected onto their initial state in the originating, stationary laboratory frame. Here it is proved that maximal entanglement is preserved in the final state. The final pure state operator is
\begin{equation}\label{B1}
\hat\rho_f = \braket{\psi_f | \psi_f },
\end{equation}
where
\begin{equation}\label{B2}
\ket{\psi} = \frac{1}{\sqrt{2}}\bigl( \ket{\alpha_f}_1 \ket{\beta_f}_2 + \ket{\beta_f}_1 \ket{\alpha_f}_2 \bigr).
\end{equation}
The form of $\ket{\alpha_f}$ and $\ket{\beta_f}$ is given in Eq. \ref{alphabeta}. When applied to Eq. \ref{B1}, the final state operator becomes
\begin{align}\label{B3}
\hat\rho_f = \frac{1}{2}\big[ &\cos(2\gamma)\bigl( \ket{R}_1\ket{T_f}_2  + \ket{T_f}_1\ket{R}_2 \bigr) \nonumber \\ 
& + \sin(2\gamma)\bigl( \ket{R}_1\ket{R}_2 - \ket{T_f}_1\ket{T_f}_2 \bigr) \bigr] \nonumber \\ 
\bigl[ &\cos(2\gamma)\bigl( \bra{R}_1\bra{T_f}_2  + \bra{T_f}_1\bra{R}_2 \bigr) \nonumber \\ 
& + \sin(2\gamma)\bigl( \bra{R}_1\bra{R}_2 - \bra{T_f}_1\bra{T_f}_2 \bigr) \bigr]
\end{align}
Now take the partial trace of this operator over all photon-2 modes to generate a reduced state operator:
\begin{equation}\label{B4}
\hat\rho_{\rm red,f} = {\rm Tr}_2\hat\rho_f = _2\!\!\braket{T_f | \psi_f | T_f}_2 +  _2\!\!\braket{R | \psi_f | R}_2.
\end{equation}
Evaluate the partial trace of the first term on the right-hand side of this equation to find that
\begin{align}\label{B5}
_2\!\braket{T_f | \psi_f | T_f}_2 = \frac{1}{2} &\bigl( \cos(2\gamma)\ket{R} - \sin(2\gamma)\ket{T_f} \bigr) \nonumber \\
*&\bigl( \cos(2\gamma) \bra{R} - \sin(2\gamma)\bra{T_f} \bigr) 
\end{align}
Here the subscript "1" has been dropped since this is understood to be the state operator of a single photon. The expression can be expanded and then further reduced to the weighted sum of four single-particle outer products:
\begin{align}\label{B6}
\hspace{-1 cm} _2\!\braket{T_f | \psi_f | T_f}_2  = \frac{1}{2} \bigl( & \nonumber \\
&\quad\cos^2(2\gamma)\ket{R}\bra{R} \nonumber \\
& -\cos(2\gamma)\sin(2\gamma)\ket{R}\bra{T_f} \nonumber \\
& -\cos(2\gamma)\sin(2\gamma)\ket{T_f}\bra{R} \nonumber \\
& + \sin^2(2\gamma)\ket{T_f}\bra{T_f}   \nonumber \\
&\bigr) .
\end{align}

Next use the same approach to evaluate the partial trace of the first term on the right-hand side of Eq. \ref{B4}.
\begin{align}\label{B7}
_2\!\braket{R | \psi_f | R}_2 = \frac{1}{2} &\bigl( \cos(2\gamma)\ket{T_f} - \sin(2\gamma)\ket{R} \bigr) \nonumber \\
*&\bigl( \cos(2\gamma) \bra{T_f} - \sin(2\gamma)\bra{R} \bigr) 
\end{align}
Here again the subscript "1" has been dropped since this is understood to be the state operator of a single photon. As for the first term in Eq. \ref{B4}, this expression can be expanded and then further reduced to the weighted sum of four single-particle outer products:
\begin{align}\label{B8}
\hspace{-1 cm} _2\!\braket{R | \psi_f | R}_2  = \frac{1}{2} \bigl( & \nonumber \\
&\quad\cos^2(2\gamma)\ket{T_f}\bra{T_f} \nonumber \\
& +\cos(2\gamma)\sin(2\gamma)\ket{T_f}\bra{R} \nonumber \\
& +\cos(2\gamma)\sin(2\gamma)\ket{R}\bra{T_f} \nonumber \\
& + \sin^2(2\gamma)\ket{R}\bra{R}   \nonumber \\
&\bigr) .
\end{align}
Finally, substitute the expressions of Eqs. \ref{B6} and \ref{B8} into Eq. \ref{B4} to find that
\begin{equation}\label{B9}
\hat\rho_{\rm red,f} = \frac{1}{2} \bigl( \braket{T_f |  T_f} +  \braket{R |  R} .
\end{equation}
By inspection, the eigenvalues, $\lambda_1$  and $\lambda_2$, of this reduced state operator are each equal to one-half, and the Schmidt number, K, is given by\cite{Preskill_1998}
\begin{equation}\label{B10}
K = \frac{1}{\lambda_1^2 + \lambda_2^2} = 2.
\end{equation}
 The final state is therefore maximally entangled.

\end{document}